\documentclass[authoryear,preprint,review,12pt]{elsarticle}

\usepackage{multirow}

\usepackage[T1]{fontenc} 
\usepackage{lmodern} 
\usepackage{graphicx} 
\usepackage{amsmath, amssymb, amsfonts} 
\usepackage{amsthm} 
\usepackage{mathrsfs} 
\usepackage{anyfontsize}
\usepackage[title]{appendix} 
\usepackage{xcolor} 
\usepackage{textcomp} 
\usepackage{manyfoot} 
\usepackage{booktabs} 
\usepackage{mathtools} 
\usepackage{float} 
\usepackage{bm} 
\usepackage[ruled,vlined]{algorithm2e}
\usepackage{algpseudocode} 

\usepackage{threeparttable} 
\usepackage{subcaption} 
\usepackage[section]{placeins} 
\usepackage{array} 
\usepackage{soul} 
\usepackage{xcolor}
\usepackage{enumitem}

\usepackage{relsize}

\begin{document}

\begin{frontmatter}



\title{Meta-Learning Neural Process for Implied Volatility Surfaces with SABR-induced Priors} 


\author[inst1]{Jirong Zhuang}

\affiliation[inst1]{organization={Department of Mathematics, University of Macau},
            addressline={Avenida da Universidade Taipa}, 
            postcode={999078}, 
            state={Macau},
            country={China}}

\author[inst1]{Xuan Wu*}


\begin{abstract}    

We treat implied volatility surface (IVS) reconstruction as a learning problem guided by two principles. First, we adopt a meta-learning view that trains across trading days to learn a procedure that maps sparse option quotes to a full IVS via conditional prediction, avoiding per-day calibration at test time. Second, we impose a structural prior via transfer learning: pre-train on SABR-generated dataset to encode geometric prior, then fine-tune on historical market dataset to align with empirical patterns. We implement both principles in a single attention-based Neural Process (Volatility Neural Process, VolNP) that produces a complete IVS from a sparse context set in one forward pass. On SPX options, the VolNP outperforms SABR, SSVI, and Gaussian process. Relative to an ablation trained only on market data, the SABR-induced prior reduces RMSE by about 40\% and suppresses large errors, with pronounced gains at long maturities where quotes are sparse. The resulting model is fast (single pass), stable (no daily recalibration), and practical for deployment at scale.

\end{abstract}



\end{frontmatter}




\section{Introduction}
Constructing the implied volatility surface (IVS) is a fundamental task in quantitative finance, essential for derivative pricing and risk management. A popular class of approaches comprises structural models (e.g., the SABR model \citep{Hagan2002}, Heston model \citep{heston1993closed}, Variance Gamma \citep{madan1998variance}, Kou model \citep{kou2002jump} and rough volatility models \citep{bayer2016pricing,gatheral2022volatility}) and related parametric models \citep{gatheral2004parsimonious, gatheral2014arbitrage}. These models typically assume that the underlying asset follows a stochastic differential equation and are effective at capturing the general shape of the volatility smile. However, their fixed mathematical forms can be overly rigid, failing to capture more complex market patterns. Data-driven regressors such as Gaussian process regression \citep{Spiegeleer2018,roberts2021probabilistic} offer flexibility but can overfit sparse and noisy quotes, producing surfaces that violate basic financial plausibility (e.g., no-arbitrage). A common limitation across both methods is to treat each trading day as an isolated fitting problem, which leads to per-day recalibration.

We therefore frame IVS construction as a meta-learning problem: by meta-learning we mean training across days to learn a single procedure for few-shot conditional prediction. Concretely, we propose the Volatility Neural Process (VolNP), a neural process model \citep{garnelo2018conditional, garnelo2018neural} that maps a small context set of option quotes to a full surface via conditional prediction. At test time, parameters are fixed (i.e., with no per-day updates), hence daily recalibration is avoided. The IVS is produced by a single forward pass given the sparse context set. However, like other data-driven methods, Neural Process is still prone to generating unstable and highly irregular surfaces in sparse market data. A growing body of work seeks to address this issue by incorporating financial knowledge into various data-driven methods, often through the imposition of constraints (e.g., no-butterfly arbitrage) derived from financial theory \citep{Ackerer2020,chataigner2021beyond,zheng2021incorporating,hoshisashi2023no,cont2023simulation,ning2023arbitrage,vuletic2024volgan,gonon2024operator}.

\begin{figure}[]
    \centering
    \includegraphics[width=\textwidth]{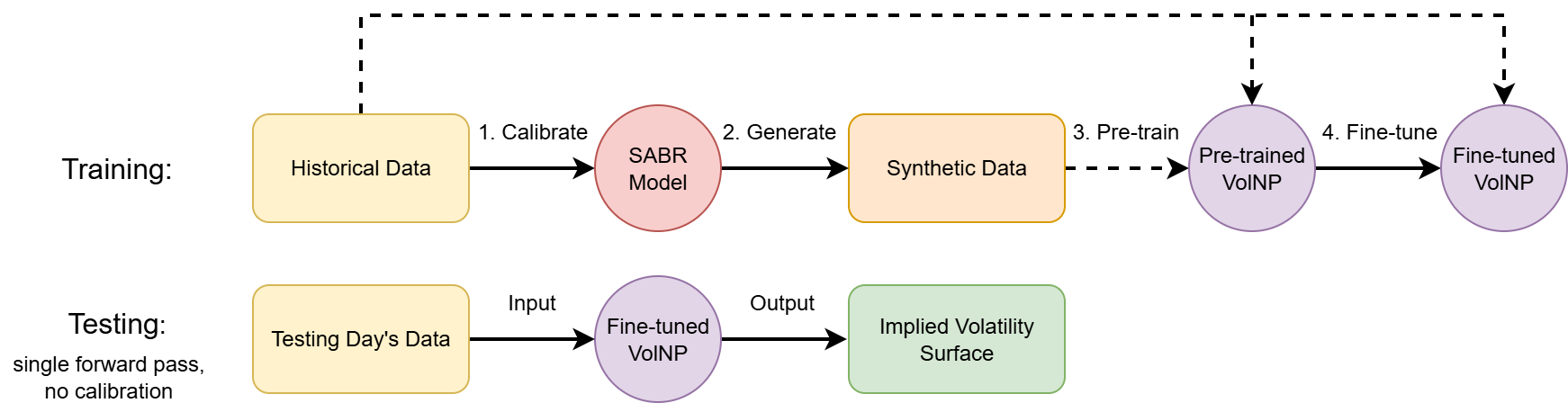}
    \caption{Flowchart for VolNP training and testing. Solid arrows show computation: historical data calibrate SABR, SABR generates synthetic IV surfaces, VolNP is pre-trained then fine-tuned, and at test time a single forward pass returns a full IV surface (no per-day calibration). Dashed arrows indicate “use as dataset” relations. In pre-training, real quotes are context and synthetic surfaces are targets; in fine-tuning, real quotes provide both context and targets.}
    \label{fig:training_flow}
\end{figure}

In contrast to these constraint-based methods, our approach integrates transfer learning \citep{chen2023teaching}. VolNP is first pre-trained on IVS generated by SABR models to encode a geometric prior for plausible volatility smiles, and is then fine-tuned on historical market data to align the prior with empirical patterns. This two-stage training (cf. Figure \ref{fig:training_flow}) reduces large errors, mitigates irregularities that arise in flexible models under sparsity, and lowers the incidence of butterfly-arbitrage violations.

\paragraph{Contributions}
Our contributions are in two aspects. First, we implement a meta-learning framework to learn a procedure that can reconstruct an implied volatility surface from a limited set of quotes, eliminating the need for recalibration. For trading desks that need to update the IVS intraday, a single forward pass enables frequent surface updates, thus removing the operational bottleneck. Second, we apply transfer learning to incorporate a structural prior from SABR pre-training, followed by market fine-tuning. This reduces large errors in sparse data scenarios, improving model accuracy and robustness. VolNP processes sparse quotes and outputs a dense surface with quantified uncertainty, making it easy to integrate into existing pricing and risk management systems.

The rest of this paper is organized as follows. Section \ref{sec2} formulates the problem. Section \ref{sec3} describes our model and training methodology. Section \ref{sec4} presents the experimental results, and Section \ref{sec6} concludes the paper.

\section{Problem Formulation}\label{sec2}

On any given trading day $q$, the implied volatility surface (IVS) can be represented by a function \(f_q: \mathbb{R}^2 \to \mathbb{R}^+\). This function maps a coordinate \(\mathbf{x} = [k, \tau]^\intercal\) (log-moneyness and time-to-expiry) to its corresponding implied volatility \(y\). We define log-moneyness as \(k=\log(K/F_q)\), where \(K\) is the strike price, and \(F_q\) is the day $q$ forward price. The function \(f_q\) is unobservable because only a finite set of market quotes is available. Let $\mathcal{Q}$ denote the index set of historical trading days used for training. Our goal is to train a single model across these days that, given a small set of quotes from day $q$, returns the conditional distribution of $y$ at any $\mathbf x$ on that day. This frames IVS construction as learning a procedure rather than fitting a new model every day.

For day $q\in\mathcal{Q}$, let $\mathcal{D}_q=\{(\mathbf{x}_j,y_j)\}_{j=1}^{N_q}$ be the finite set of $N_q$ observed option quotes. During training, we form two disjoint subsets from $\mathcal{D}_q$. The context set $\mathcal{D}_C\subset\mathcal{D}_q$ is the information that the model is allowed to condition on for that day. The target set $\mathcal{D}_T\subset\mathcal{D}_q$ is used only to define the loss and is not given to the model when making prediction. The learning objective is to train a model $p_\theta(y\mid \mathbf{x},\mathcal{D}_C)$ that maps the day’s sparse context quotes to a predictive distribution at any coordinate $\mathbf{x}$, where $\theta$ denotes parameters in the model. This frames IVS construction as a set-to-function regression problem: the input is a set $\mathcal{D}_C$ and the output is a function on $(k,\tau)$. 

In this paper, we introduce the Volatility Neural Process (VolNP) to model the conditional distribution $p_\theta(y\mid \mathbf{x},\mathcal{D}_C)$. By training across many days in $\mathcal{Q}$, the VolNP learns how to use a few quotes to reconstruct a full surface. In deployment on a new day $q^\prime\notin\mathcal{Q}$, we often need to construct an IVS with sparse intraday quotes. The workflow is direct: use the currently available quotes as the context set $\mathcal{D}_C$, and evaluate the conditional distribution on any grid of $\mathbf{x}$. The model parameters remain fixed during deployment; there is no daily calibration or gradient update. A single forward pass produces a complete IVS (and associated uncertainty) for option pricing and risk management.

\section{Details of Volatility Neural Process}\label{sec3}

\subsection{Model Architecture}
To address the set-to-function regression task defined within our meta-learning framework, we propose the Volatility Neural Process (VolNP), a variant of the Attentive Neural Process \citep{kim2019attentive}. We implement meta-learning and transfer learning as follows: meta-learning learns a procedure across days, while transfer learning injects a structural prior via SABR pre-training and adapts it via market fine-tuning. The following describes how the model processes the inputs for a single task. Such a task consists of a context set \(\mathcal{D}_{C}\) and a set of target coordinates \(\mathcal{X}_{T}\). For notational simplicity, we will use these symbols throughout this subsection to refer to the inputs of a single forward pass of the model.

\begin{figure}[]
    \centering
    \includegraphics[width=0.85\textwidth]{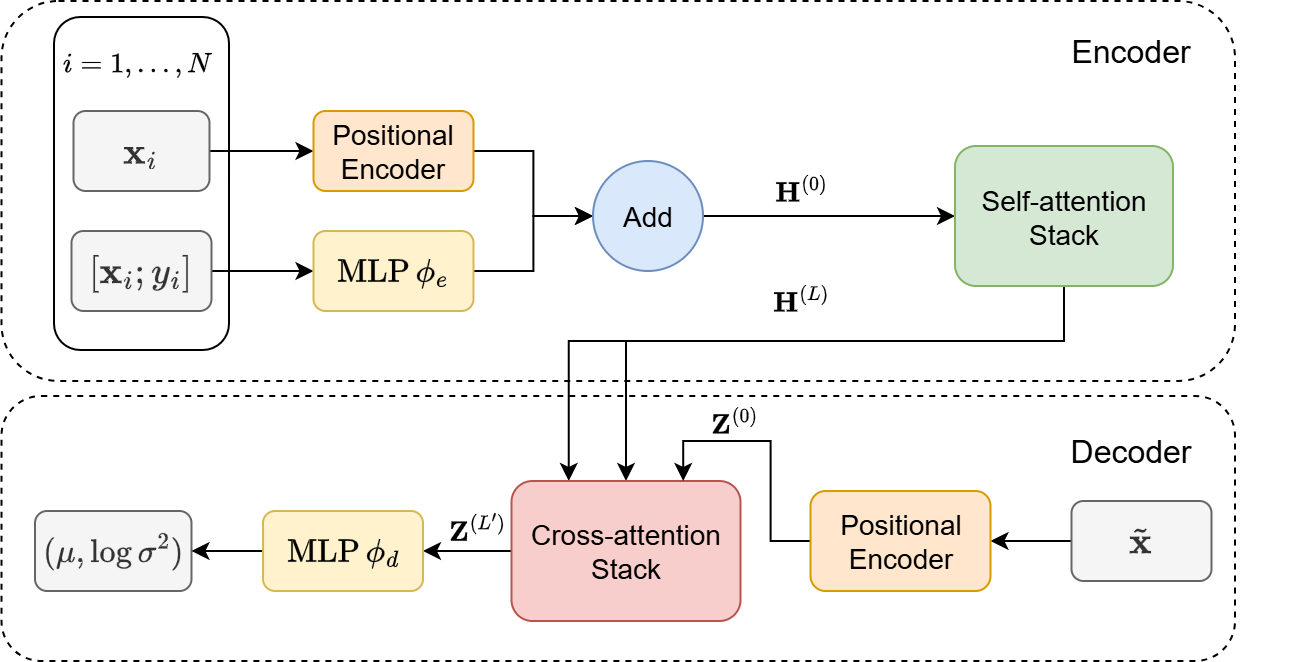}
    \caption{The architecture of the Volatility Neural Process.}
\label{volnp}
\end{figure}
The model is an encoder-decoder architecture (cf. Figure \ref{volnp}) with attention mechanisms, designed to effectively model the complex interactions within a set of context points and reason about target predictions.

\subsubsection{Encoder}
The encoder's goal is to learn the complex relationships between the individual option quotes in the context set \(\mathcal{D}_C\). It maps this set to a matrix of contextualized representations, \(\mathbf{H}^{(L)} \in \mathbb{R}^{N \times d_r}\), where \(d_r\) is the hidden dimension, and each row contains a summary of a single point informed by all other points through the self-attention mechanism. Here, \(N\) denotes the number of data points in \(\mathcal{D}_C\), and \(L\) indicates the number of self-attention blocks.

The process begins by creating an initial representation \(\mathbf{h}_i^{(0)}\) for each context point \((\mathbf{x}_i, y_i)\):
\begin{equation}
    \mathbf{h}_i^{(0)} = \phi_e([\mathbf{x}_i; y_i]) + \gamma(\mathbf{x}_i)
    \label{eq:initial_repr}
\end{equation}
where \(\phi_e: \mathbb{R}^3 \to \mathbb{R}^{d_r}\) is a learnable Multilayer Perceptron (MLP) and \([\cdot; \cdot]\) denotes concatenation. This step combines two types of information. First, \(\phi_e\) extracts a set of features from the raw data. Second, we add a fixed sinusoidal positional encoding \citep{vaswani2017attention}, \(\gamma(\mathbf{x}_i)\). The resulting representation \(\mathbf{h}_i^{(0)}\) thus contains information about both the specific value of an option quote and its absolute position on the moneyness-expiry grid. The collection of initial states forms the matrix \(\mathbf{H}^{(0)} = \left[\mathbf{h}_1^{(0)}, \dots, \mathbf{h}_{N}^{(0)}\right]^\intercal\).

Next, these initial representations are processed by a stack of \(L\) self-attention blocks (\(\text{SA}\)). This is the core mechanism that allows the model to reason about the global structure of the volatility surface. A single option quote is only meaningful in the context of others. For instance, the implied volatility of an at-the-money option is intrinsically linked to its out-of-the-money counterparts, collectively forming the "smile" for a given maturity. The self-attention mechanism formalizes this intuition. It allows each point \(\mathbf{h}_i^{(0)}\) to query all other points on the surface and update its representation based on their relationships. Through this process, the model learns to identify and encode complex, non-local patterns directly from the sparse data.
\begin{equation}
    \mathbf{H}^{(l)} = \text{SA}(\mathbf{H}^{(l-1)}) \quad \text{for } l = 1, \dots, L
    \label{eq:contextualization}
\end{equation}
Each SA block is built around a Multi-Head Attention (MHA) mechanism \citep{vaswani2017attention} and uses a Pre-LayerNorm architecture \citep{xiong2020layer} for better training stability. 

In this self-attention step, the Query (\(\mathbf{Q}\)), Key (\(\mathbf{K}\)), and Value (\(\mathbf{V}\)) matrices are all derived from the same input, \(\mathbf{H}^{(l-1)}\). The model compares each option quote (Query) to all other quotes (Key) to decide how much information to draw from them (Value). The model performs this comparison not just once, but with \(n_h\) parallel "attention heads". Here, the projection matrices \(\mathbf{W}_i^\mathbf{Q}, \mathbf{W}_i^\mathbf{K} \in \mathbb{R}^{d_r \times d_k}\), \(\mathbf{W}_i^\mathbf{V} \in \mathbb{R}^{d_r \times d_v}\), and \(\mathbf{W} \in \mathbb{R}^{n_h d_v \times d_r}\) are learnable parameters, with \(d_k = d_v = d_r / n_h\). Each head can learn to focus on different types of financial relationships. The output of the $i$-th head \(\mathbf{a}_i\) is:
\begin{equation}
    \mathbf{a}_i = \text{softmax}\left(\frac{(\mathbf{Q}\mathbf{W}_i^\mathbf{Q})(\mathbf{K}\mathbf{W}_i^\mathbf{K})^\intercal}{\sqrt{d_k}}\right) (\mathbf{V}\mathbf{W}_i^\mathbf{V})
\end{equation}
The outputs of all heads are then combined to produce the final MHA output:
\begin{equation}
    \text{MHA}(\mathbf{Q},\mathbf{K}, \mathbf{V}) =[\mathbf{a}_1\, ; \dots\, ; \mathbf{a}_{n_h}] \mathbf{W}
\end{equation}
The full SA block uses a Pre-LayerNorm architecture for better training stability:
\begin{align}
    \mathbf{H}' &= \mathbf{H}^{(l-1)} + \text{MHA}(\text{LN}(\mathbf{H}^{(l-1)}), \text{LN}(\mathbf{H}^{(l-1)}), \text{LN}(\mathbf{H}^{(l-1)})) \\
    \mathbf{H}^{(l)} &= \mathbf{H}' + \phi_{\text{ffn}}(\text{LN}(\mathbf{H}'))
\end{align}
where \(\text{LN}\) is Layer Normalization and \(\phi_{\text{ffn}}\) is a feed-forward network.

After \(L\) layers, the final output \(\mathbf{H}^{(L)}\) contains representations where each point is informed by the global market structure of that day.

\subsubsection{Decoder}
The decoder uses the contextualized representation \(\mathbf{H}^{(L)}\) to predict the output distribution for any target coordinate \(\tilde{\mathbf{x}}_j \in \mathcal{X}_T\). This is achieved by treating the decoding process as a query mechanism. First, each target coordinate is mapped to an initial query vector using the same positional encoding function:
\begin{equation}
    \mathbf{z}_j^{(0)} = \gamma(\tilde{\mathbf{x}}_j)
    \label{eq:initial_target_repr}
\end{equation}
These query vectors, forming the matrix \(\mathbf{Z}^{(0)}=\left[\mathbf{z}_1^{(0)},\dots,\mathbf{z}_M^{(0)}\right]^\intercal\), are then refined through a stack of \(L^\prime\) cross-attention blocks (\(\text{CA}\)). The decoding process can be interpreted as using a target's location to query the rich, contextualized information database built from the observed data points. This allows each target point to aggregate the most relevant information for its own prediction. For each target, it computes a set of relevance scores against all points in the context set, effectively learning a unique set of weights for combining the context representations.
\begin{equation}
    \mathbf{Z}^{(l)} = \text{CA}(\mathbf{Z}^{(l-1)}, \mathbf{H}^{(L)}) \quad \text{for } l = 1, \dots, L^\prime
    \label{eq:cross_attention}
\end{equation}
The CA block uses the MHA, where queries come from the target representations \(\mathbf{Z}\) and the Keys/Values come from the context representation \(\mathbf{H}^{(L)}\).
The final target representations, \(\mathbf{Z}^{(L^\prime)} = \left[\mathbf{z}_1^{(L^\prime)}, \dots, \mathbf{z}_{M}^{(L^\prime)}\right]^\intercal\), which contains all necessary information distilled from the context set for each specific target, are passed through a MLP, \(\phi_d: \mathbb{R}^{d_r} \to \mathbb{R}^2\):
\begin{equation}
    [\mu_j, \log\sigma_j^2]^\intercal = \phi_d(\mathbf{z}_j^{(L^\prime)})
    \label{eq:prediction_head}
\end{equation}
This yields the final predictive distribution \(\mathcal{N}(\mu_j, \sigma_j^2)\) for the volatility at coordinate \(\tilde{\mathbf{x}}_j\).

\subsection{Training}
Our model is trained using an approach comprising two stages: pre-training provides a geometric prior; fine-tuning adapts this prior to real market data. The training process relies on two datasets.

\subsubsection{Datasets and Task Generation}

The first dataset is the collection of historical market data, \(\{\mathcal{D}_q\}_{q \in \mathcal{Q}}\), where \(\mathcal{D}_q = \{(\mathbf{x}_j, y_j)\}_{j=1}^{N_q}\) is the set of all \(N_q\) observed option quotes for a given trading day \(q\). The second is a synthetic dataset, \(\{\hat{\mathcal{D}}_q\}_{q \in \mathcal{Q}}\), where \(\hat{\mathcal{D}}_q\) is the corresponding dense, smooth volatility surface generated by a calibrated SABR model. The SABR model is calibrated to each maturity slice in $\mathcal{D}_q$, and parameters are linearly interpolated across the term structure.

At each training step, a new task is generated. This process begins by randomly sampling a day index \(q\) from the set of available historical trading days \(\mathcal{Q}\). Subsequently, a context set \(\mathcal{D}_{C}\) and a target set \(\mathcal{D}_{T}\) are constructed by drawing samples from the data sources \(\mathcal{D}_q\) or \(\hat{\mathcal{D}}_q\), with the specific sampling strategy depending on the training stage.

\subsubsection{Objective Function}
The model's parameters \(\theta\) are optimized by minimizing the negative log-likelihood (NLL) over the target sets. For a single task generated from day \(q\), the loss is defined over its context set \(\mathcal{D}_C\) and target set \(\mathcal{D}_T = \{(\tilde{\mathbf{x}}_j, \tilde{y}_j)\}_{j=1}^M\):
\begin{align*}
       \mathcal{L}(\theta; \mathcal{D}_C, \mathcal{D}_T) &= -\sum_{j=1}^{M} \log p_\theta(\tilde{y}_j \mid \mathcal{D}_C, \tilde{\mathbf{x}}_j) \\
       &= \frac{1}{2} \sum_{j=1}^{M} \left( \frac{(\tilde{y}_j - \mu_j)^2}{\sigma_j^2} + \log\sigma_j^2 \right).
\end{align*}

In each training step, we sample a task and compute the gradient of this loss function.

\subsubsection{Two-stage Training}

The training curriculum is a two-stage process designed to first instill a robust prior and then adapt it to market specifics (cf. Figure \ref{fig:training_flow}). 

\begin{itemize}
   \item \textbf{Stage 1: Pre-training.} The goal is to instill a structural, financial-theoretic prior. For each task, a day \(q\) is sampled, and the context set \(\mathcal{D}_{C}\) is drawn from the sparse, real market quotes \(\mathcal{D}_q\), while the target set \(\hat{\mathcal{D}}_{T}\) is drawn from the corresponding dense, synthetic SABR surface \(\hat{\mathcal{D}}_q\). The model's parameters \(\theta\) are then updated by minimizing the loss \(\mathcal{L}(\theta; \mathcal{D}_{C}, \hat{\mathcal{D}}_{T})\), encoding geometric plausibility of volatility smiles.

   \item \textbf{Stage 2: Fine-tuning.} This stage adapts the prior to real-world market dynamics. The model is initialized with the pre-trained parameters and, for each task, both the context set \(\mathcal{D}_{C}\) and the target set \(\mathcal{D}_{T}\) are sampled exclusively from the real market data \(\mathcal{D}_q\), ensuring the sets are non-overlapping. The optimization objective remains \(\mathcal{L}(\theta; \mathcal{D}_{C}, \mathcal{D}_{T})\).
\end{itemize}

\section{Experiments}
\label{sec4}

\subsection{Experimental Setup}
Our experiments are carried out on a dataset of S\&P 500 (SPX) index options, spanning from January 2006 to August 2023. The raw data is from \textit{OptionMetrics} dataset and is strictly partitioned by time to ensure a fair evaluation. Data from 2006 to 2018 is used for training and validation, with 10\% of the trading days randomly held out for validation. Data from 2019 to 2023 serves as the out-of-sample test set for all final evaluations. Quotes are preprocessed by computing mid-prices from best bid/ask and applying liquidity filters: positive bid and ask with bid$<$ask; daily volume $>5$; relative bid-ask spread $\le 50\%$; and mid-price $\ge 0.1$. Forward prices are computed from the zero-coupon curve and dividend yields (from \textit{OptionMetrics}), and implied volatilities are obtained under the Black framework. Descriptive statistics of the dataset are in Table~\ref{table:dataset_stats}.

\begin{table}[ht]
\centering
\caption{Descriptive statistics of the dataset. The dataset is partitioned by time into a training set and a test set.}
\label{table:dataset_stats}
\resizebox{0.85\textwidth}{!}{%
\begin{tabular}{llrr}
\toprule
\textbf{Statistic} & \textbf{Category} & \textbf{Training Set} & \textbf{Test Set} \\
\midrule

\multirow{3}{*}{Maturity} & Short-Term ($\le$ 3M) & 1,355,448 (75.0\%) & 1,723,463 (70.5\%) \\
 & Mid-Term (3M-1Y) & 367,168 (20.3\%) & 633,933 (25.9\%) \\
 & Long-Term ($>$ 1Y) & 84,265 (4.7\%) & 88,796 (3.6\%) \\
\midrule
\multirow{3}{*}{Moneyness} & $|k| \le 0.05$ & 903,275 (50.0\%) & 1,183,639 (48.4\%) \\
 & $0.05 < |k|  \le 0.2$ & 747,269 (41.4\%) & 1,047,066 (42.8\%) \\
 & $|k| > 0.2$ & 156,337 (8.7\%) & 215,487 (8.8\%) \\
\bottomrule
\end{tabular}%
}
\end{table}

We evaluate our proposed \textbf{VolNP-SP} \footnote{`SP' stands for `SABR-induced Prior'. The hidden dimension of VolNP models \(d_r\) is 128. Both the encoder and decoder MLPs (\(\phi_e, \phi_d\)) consist of 3 hidden layers with 128 units each and GELU activation. We use \(L=3\) self-attention blocks and \(L^\prime=3\) cross-attention blocks, each with 4 attention heads.} against an ablation version, \textbf{VolNP-Base}\footnote{All VolNP models are trained with the AdamW optimizer. VolNP-SP is pre-trained for up to 200 epochs with a learning rate of \(5 \times 10^{-5}\) and fine-tuned for up to 200 epochs with a learning rate of \(2 \times 10^{-6}\). The VolNP-Base model is trained for up to 400 epochs with a learning rate of \(5 \times 10^{-5}\).} (trained only on market data), and three benchmarks. The benchmarks are: (1) a slice-calibrated \textbf{SABR} model\footnote{SABR is calibrated to each maturity slice, and its parameters are linearly interpolated across the term structure.}, (2) a \textbf{Gaussian process (GP)} with an RBF kernel, and (3) a Surface SVI (\textbf{SSVI}) model\footnote{The SSVI model uses the arbitrage-free parameterization of \citet{gatheral2014arbitrage}. }.  For each test day, we uniformly sample a context set of \(N=100\) quotes from all available quotes of that day; the target set comprises the remaining quotes for that day. All baselines and VolNP variants are evaluated on the identical target sets. Performance is measured by Root Mean Squared Error (RMSE), Mean Absolute Error (MAE), and coefficient of determination (R$^2$). Let $(y_j, \hat y_j)_{j=1}^M$ denote true and predicted implied volatilities and let $\bar y = \tfrac{1}{M}\sum_{j=1}^M y_j$,
\begin{align*}
    \mathrm{MAE} =& \frac{1}{M} \sum_{j=1}^M \big| y_j - \hat y_j \big|\\
    \mathrm{RMSE} =& \sqrt{\frac{1}{M} \sum_{j=1}^M \big( y_j - \hat y_j \big)^2}\\
    R^2 =& 1 - \frac{\sum_{j=1}^M \big( y_j - \hat y_j \big)^2}{\sum_{j=1}^M \big( y_j - \bar y \big)^2}
\end{align*}
RMSE and MAE are reported in basis points (BPS), where 1 BPS equals 0.01\% in implied volatility.

\subsection{Overall and Stratified Performance}
\label{sec:performance}

\begin{table}[ht]
\centering
\caption{Overall and stratified performance comparison. Short-, mid-, and long-term correspond to maturities of 0-3 months, 3-12 months, and >12 months, respectively. RMSE and MAE are in basis points (BPS). Best result in each category is in \textbf{bold}.}
\label{table:performance}
\resizebox{1.0\textwidth}{!}{%
\begin{tabular}{lcccccccccccc}
\toprule
\multirow{2}{*}{\textbf{Model}} & \multicolumn{3}{c}{\textbf{Overall}} & \multicolumn{3}{c}{\textbf{Short-Term}} & \multicolumn{3}{c}{\textbf{Mid-Term}} & \multicolumn{3}{c}{\textbf{Long-Term}} \\
\cmidrule(lr){2-4} \cmidrule(lr){5-7} \cmidrule(lr){8-10} \cmidrule(lr){11-13}
& RMSE & MAE & R$^2$ & RMSE & MAE & R$^2$ & RMSE & MAE & R$^2$ & RMSE & MAE & R$^2$ \\
\midrule
VolNP-SP         & \textbf{96.04} & \textbf{48.22} & \textbf{0.98} & 104.03 & 50.44 & 0.98 & \textbf{71.80} & \textbf{41.48} & \textbf{0.99} & \textbf{85.44} & \textbf{53.33} & \textbf{0.98} \\
VolNP-Base       & 164.75 & 48.98 & 0.96 & 171.29 & 43.72 & 0.96 & 151.41 & 59.01 & 0.96 & 121.06 & 79.14 & 0.95 \\
SABR             & 153.71 & 67.59 & 0.96 & 135.45 & 59.83 & 0.97 & 158.84 & 70.95 & 0.95 & 338.48 & 194.22 & 0.63 \\
GP & 274.90 & 73.84 & 0.88 & \textbf{58.53} & \textbf{32.62} & \textbf{0.99} & 255.10 & 78.69 & 0.88 & 1244.98 & 838.73 & -3.94 \\
SSVI             & 268.39 & 151.34 & 0.88 & 253.84 & 143.78 & 0.90 & 269.31 & 144.51 & 0.86 & 464.59 & 346.66 & 0.31 \\
\bottomrule
\end{tabular}%
}
\end{table}

Table \ref{table:performance} presents the errors across all trading days. Overall, VolNP-SP achieves the best performance, with the lowest RMSE (96.04 BPS) and MAE (48.22 BPS), and the highest R$^2$ (0.98). A key finding comes from the comparison with VolNP-Base. While their overall MAE is similar, the pre-training strategy allows VolNP-SP to reduce the RMSE by nearly 40\%. This suggests that the SABR-induced prior does not just improve the average fit, but is particularly effective at preventing large, financially significant prediction errors.

The stratified results provide further insight. For short-term options, where market data is most dense, the flexible Gaussian process model performs best. However, as maturity increases and data becomes sparser, the performance of GP and other benchmarks degrades. In these mid- and long-term regimes, VolNP-SP outperforms all other models. This indicates that the learned geometric prior is crucial for constructing a coherent and robust surface, especially in regions with fewer data points. 

We complement the quantitative metrics with two heatmaps that aggregate errors over the full $(k,\tau)$ domain across all test days. For each test day we bin target coordinates $(k,\tau)$ on a fixed global grid. Within each bin we compute the per-day RMSE and the per-day mean signed error (bias). We then take the median across test days in each bin. 

\begin{figure}[]
    \centering
    \includegraphics[width=\textwidth]{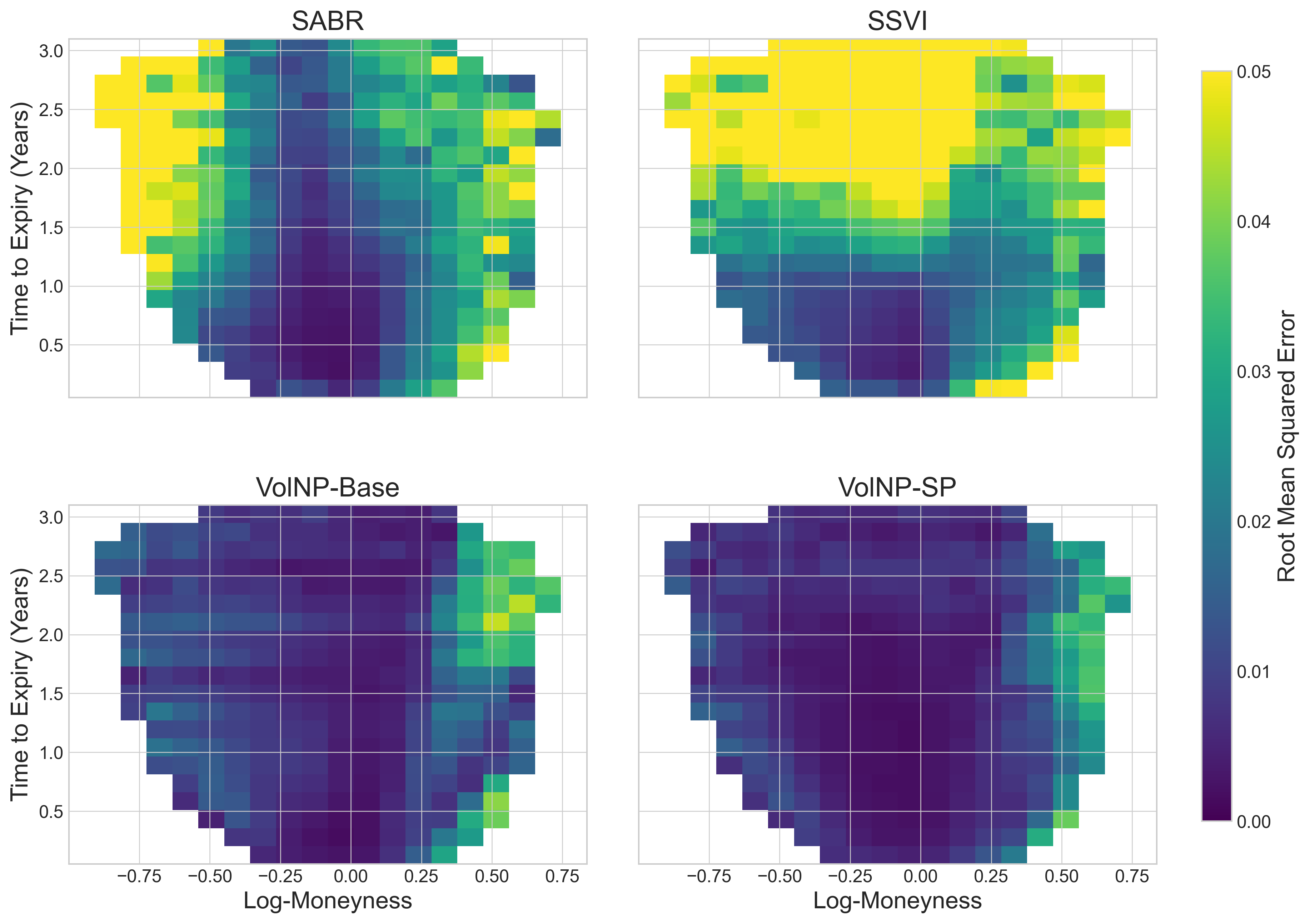}
    \caption{Root Mean Squared Error (RMSE) heatmaps for benchmark and VolNP models. Darker colors represent lower error. }
    \label{fig:error_heatmap}
\end{figure}

\begin{figure}[ht]
    \centering
    \includegraphics[width=\textwidth]{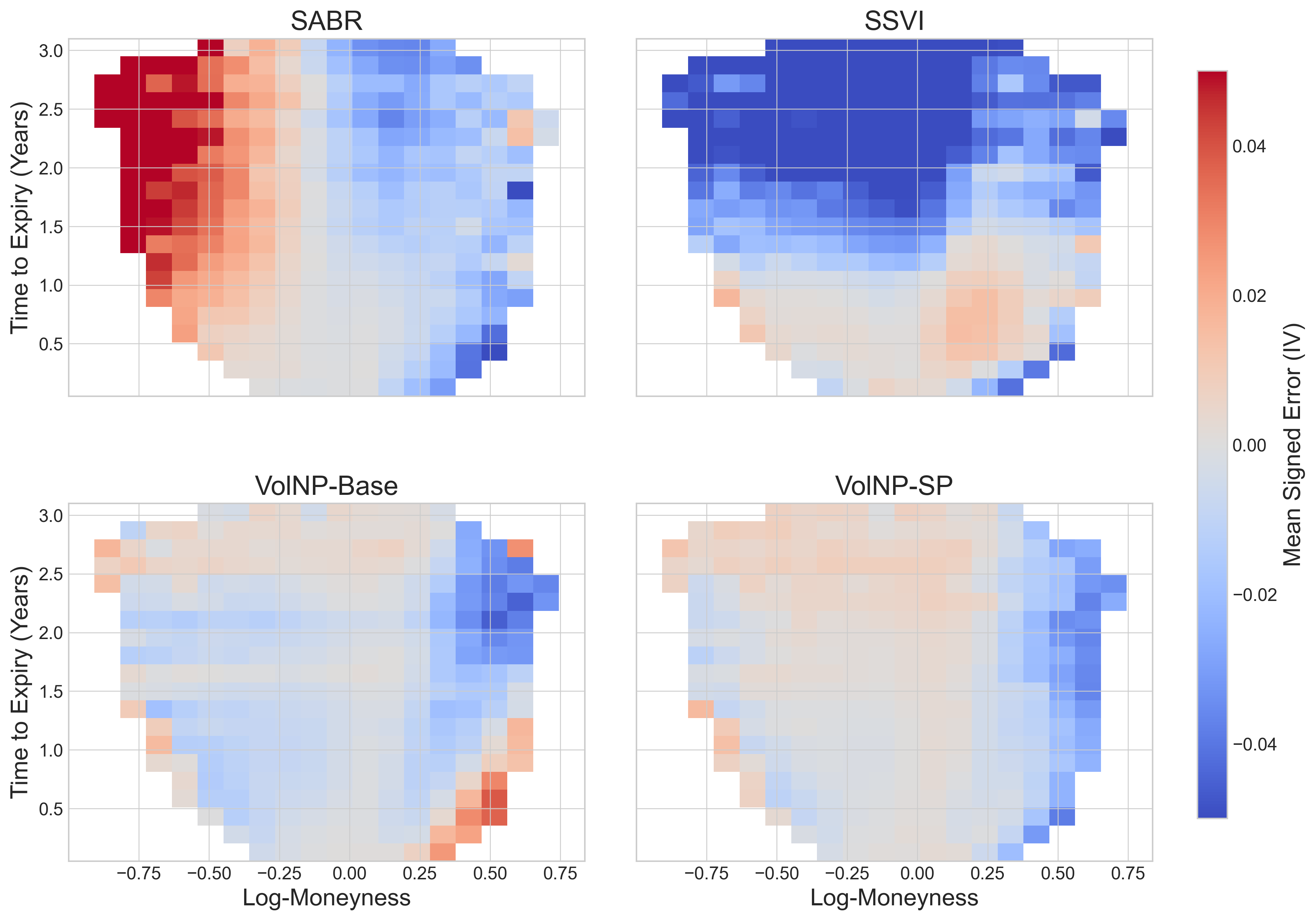}
    \caption{Mean signed error (bias) heatmaps for benchmark and VolNP models. Blue indicates underestimation; red indicates overestimation.}
    \label{fig:bias_heatmap}
\end{figure}

The RMSE heatmaps (Figure \ref{fig:error_heatmap}) confirm that VolNP-SP maintains low error not only near-the-money but also across sparse long-maturity regions where benchmarks perform poorly. The bias heatmaps (Figure \ref{fig:bias_heatmap}) show SABR and SSVI display pronounced directional biases on wings or long maturities, while VolNP-SP's bias is near-neutral and spatially smooth, consistent with its higher overall R$^2$ and lower RMSE.

\subsection{Monthly RMSE Time Series}
\label{sec:monthly_rmse}
To examine temporal robustness, we aggregate daily RMSE into monthly averages. This view highlights regime shifts and macro stress episodes, and contrasts the stability of VolNP-SP against the market-only baseline.

\begin{figure}[ht]
    \centering
    \includegraphics[width=\textwidth]{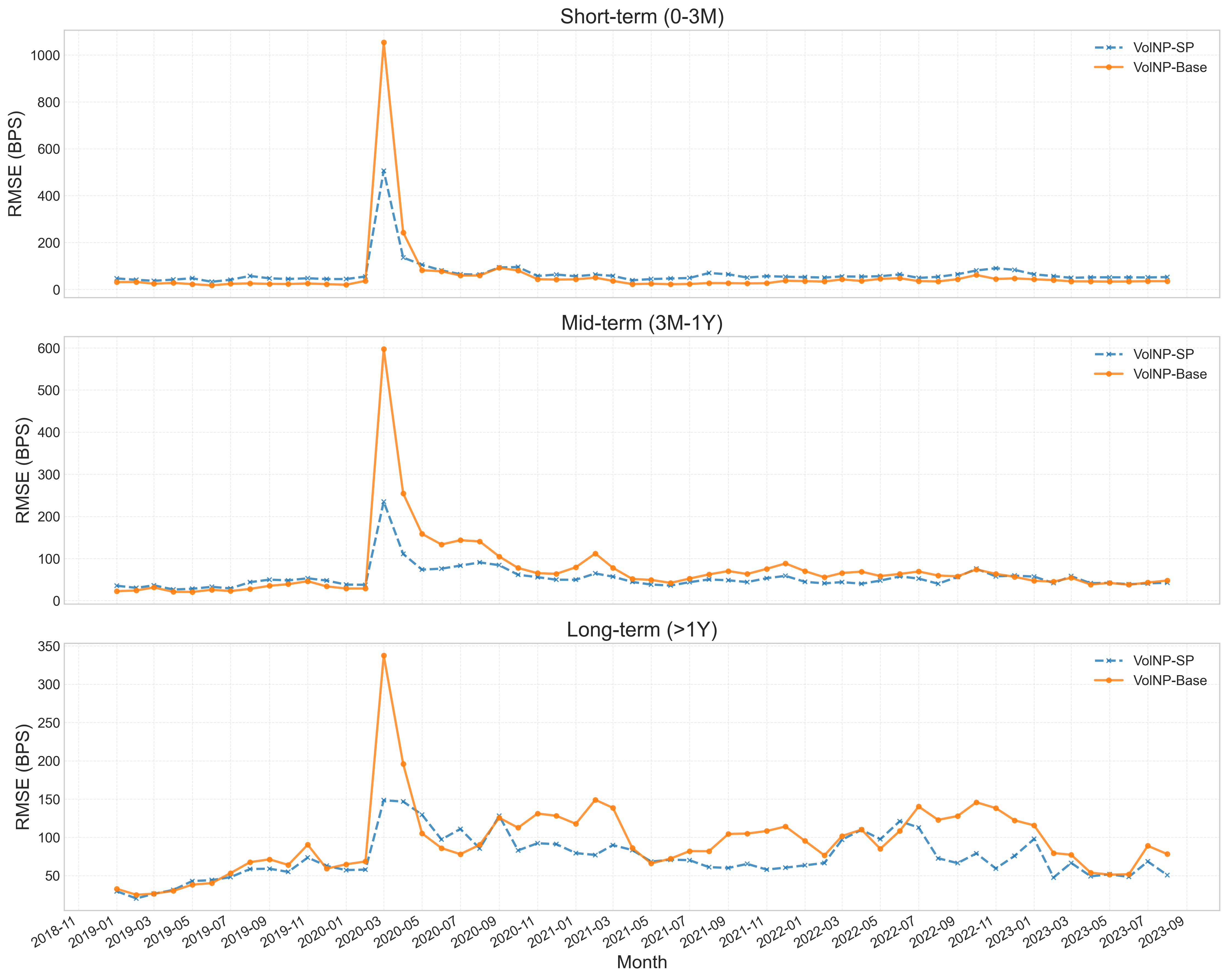}
    \caption{Monthly averaged RMSE (in BPS) over time, stratified by maturity. Each subplot compares VolNP-SP (blue) and VolNP-Base (orange).}
    \label{fig:monthly_rmse_timeseries}
\end{figure}

Figure \ref{fig:monthly_rmse_timeseries} reveals three patterns. First, during the early-2020 volatility spike, both models suffer higher errors, but VolNP-SP rises less across all maturities. Second, from 2021 onward, VolNP-SP maintains a persistent gap over VolNP-Base in mid-term and long-term, aligning with its superior in stratified metrics. Third, short-dated months exhibit small fluctuations where both models are close, but VolNP-SP remains at least as accurate.

\subsection{Computational Efficiency}
\label{sec:compute}
We report the average runtime per-day evaluation\footnote{\textit{Environment.} Python 3.10; Intel i7-12700 (2.10 GHz); NVIDIA RTX 3060 (12 GB). VolNP models run in PyTorch (CUDA) and we time a single forward pass; SABR/SSVI/GP run on CPU (NumPy/SciPy; scikit-learn for GP) and include calibration/fit plus prediction.}. One evaluation corresponds to selecting \(N=100\) context quotes for a test day and producing predictions for all remaining target quotes of that day in a single run. Table~\ref{table:comp_time} lists the mean time per evaluation in milliseconds.

\begin{table}[ht]
\centering
\caption{Average per-day evaluation runtime (milliseconds).}
\label{table:comp_time}
\resizebox{0.85\textwidth}{!}{%
\begin{threeparttable}
\begin{tabular}{l c c c c c}
\toprule
 & VolNP-Base & VolNP-SP & SSVI & GP & SABR \\
\midrule
Avg time (ms) & 23.9 & 24.5 & 123.4 & 189.6 & 1211.1 \\
\bottomrule
\end{tabular}
\end{threeparttable}}
\end{table}

Both VolNP variants require about 25 ms per evaluation, and outperform other methods. This reflects deployment cost for generating a day's IVS from sparse context, and VolNP scales well for intraday updates.

\subsection{Sensitivity to Data Sparsity}
\label{sec:sensitivity}

\begin{figure}[ht]
    \centering
    \includegraphics[width=\textwidth]{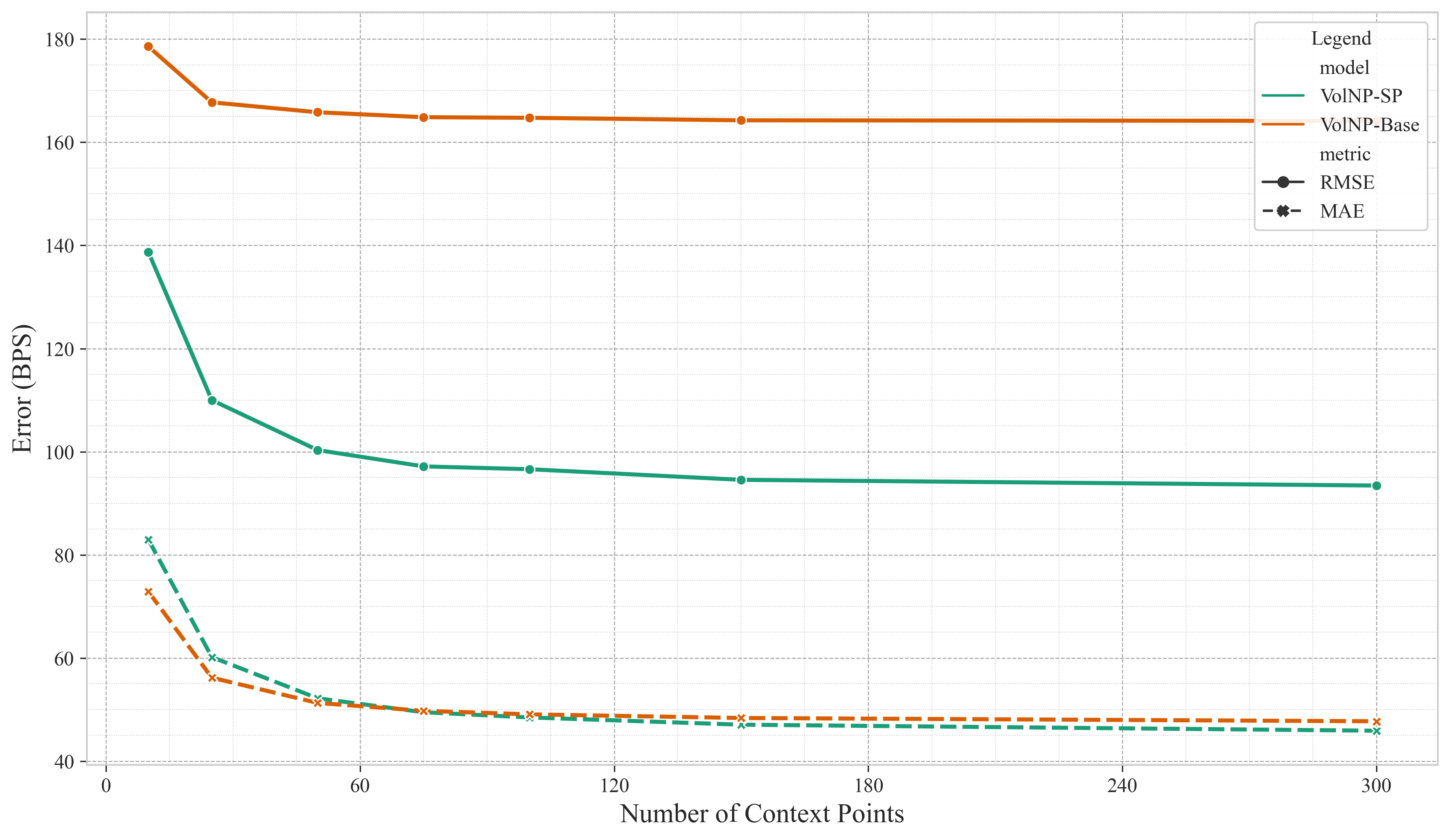}
    \caption{Model error (RMSE and MAE) as a function of the number of context points. VolNP-SP demonstrates superior data efficiency, achieving lower error with fewer context points and showing more stable performance as data becomes sparse.}
    \label{fig:sensitivity_analysis}
\end{figure}

To study the robustness of the model, we examine the performance of the model under variety data sparsity (Figure \ref{fig:sensitivity_analysis}). The RMSE metric shows that VolNP-SP maintains a lower RMSE across all levels of sparsity, demonstrating data efficiency. VolNP-SP's lower RMSE confirms its ability to maintain a globally plausible surface shape instead of just fitting local points. This highlights its reliability for practical applications.

\subsection{Case Study: Surface Reconstruction Under Market Stress}

\begin{figure}[ht]
    \centering
    \includegraphics[width=\textwidth]{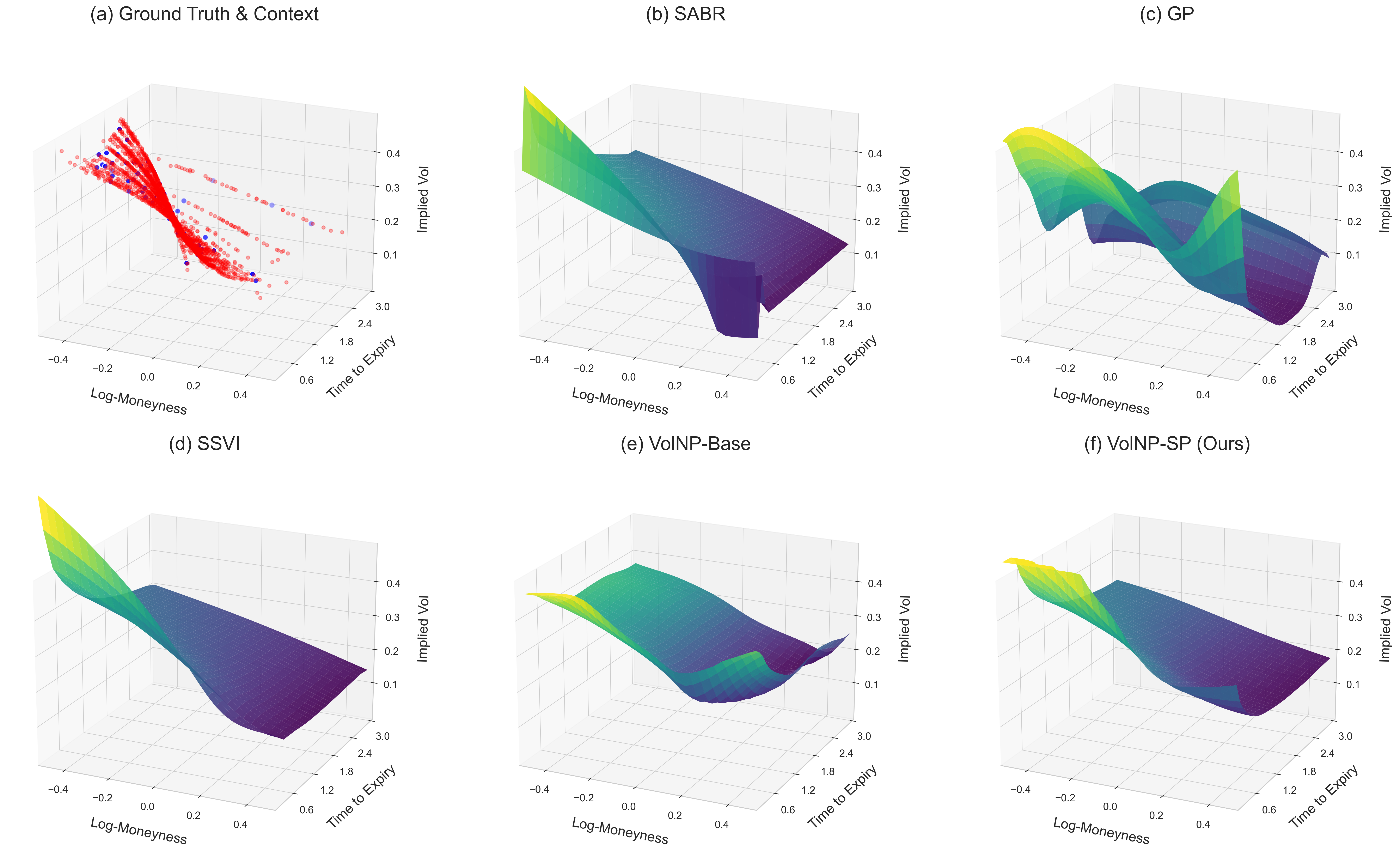}
    \caption{Implied volatility surface construction on March 9, 2020. Each model attempts to reconstruct the full surface from the same context dots (blue dots) shown in (a). Red dots in (a) are unobserved data to represent the ground truth. }
    \label{fig:surface_reconstruction}
\end{figure}

We visualize the reconstructed surfaces on a challenging day: March 9, 2020, during the COVID-19 volatility shock. This event provides a real-world stress test for model robustness. Figure \ref{fig:surface_reconstruction} provides a comprehensive comparison, where each model reconstructs the surface from the same 100 context points drawn from the market data (plotted as blue dots in panel a). Red dots in (a) are unobserved data to represent the ground truth.

The benchmark models exhibit distinct failure modes. The SSVI (d) and SABR (b) shows clear volatility smiles but are overly rigid, failing to capture finer market details. The highly flexible Gaussian process (c) overfits the sparse data, producing an unstable and financially implausible surface with unnatural oscillations. The ablation model VolNP-Base (e), while capturing some features of smiles, yields an irregular surface, suggesting it overfits local noise without a global structural guide. In contrast, VolNP-SP (f) reconstructs a surface that successfully captures the essential features of the ground truth market data. This visual evidence reinforces the quantitative results, suggesting that the geometric prior learned during pre-training is critical for regularizing the model and enabling it to generate coherent IVS from sparse dataset.

\begin{figure}[ht]
    \centering
    \includegraphics[width=\textwidth]{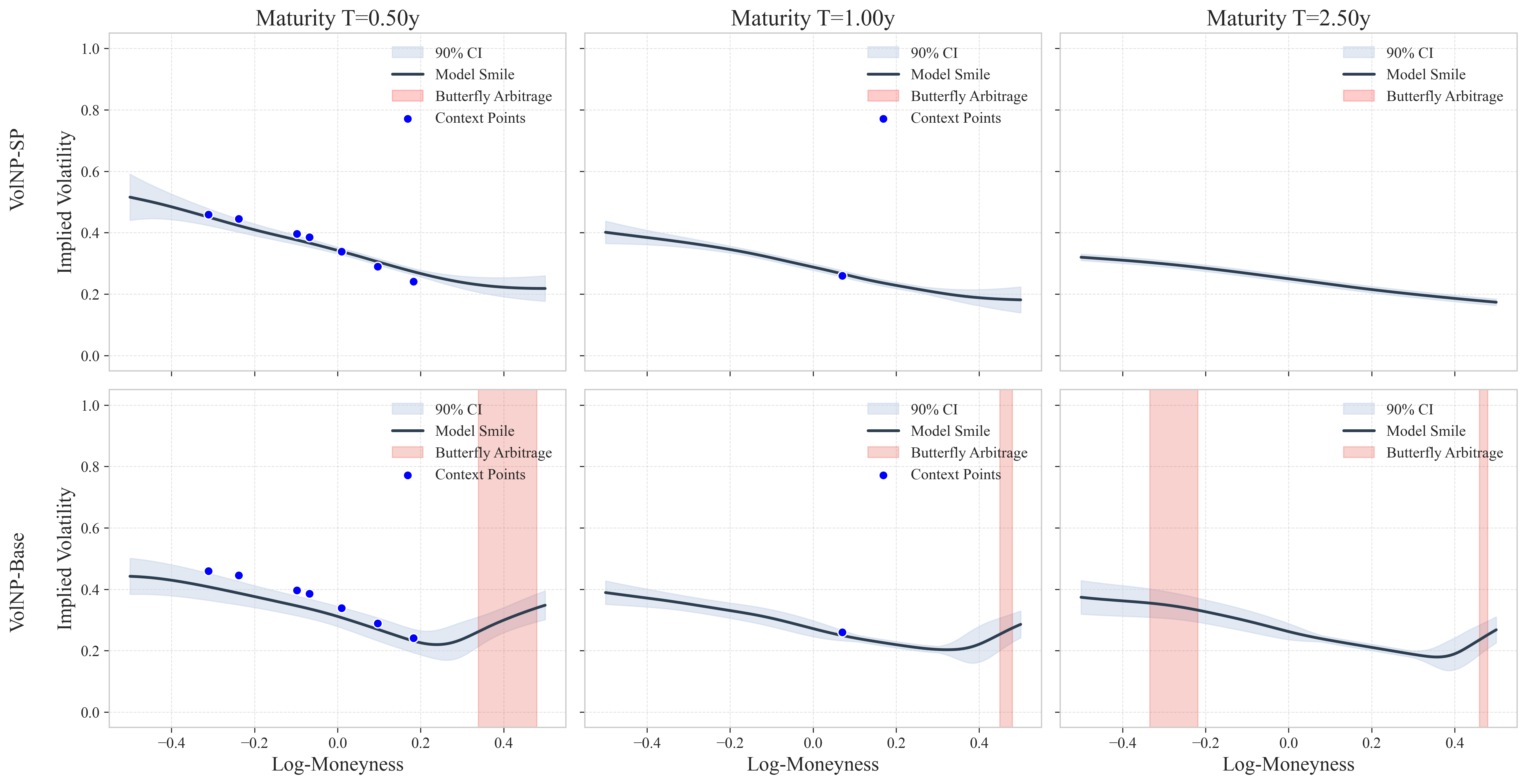}
    \caption{Arbitrage analysis on March 9, 2020. The red shaded areas indicate regions where the smile admits butterfly arbitrage (against Durrleman condition).}
    \label{fig:smile_arbitrage}
\end{figure}

 We test for violations of the no-butterfly-arbitrage condition, which ensures the model's implied risk-neutral probability density is non-negative. This test is performed by checking the Durrleman condition (further details, cf. Theorem 2.9, condition IV3 of \citep{roper2010arbitrage}). Figure \ref{fig:smile_arbitrage} shows the result on March 9, 2020. Smiles from VolNP-Base (bottom row) exhibit significant violations of this condition, indicated by the red shaded areas where arbitrage opportunities exist. This makes the model unusable in practice. In contrast, smiles from VolNP-SP (top row) are consistently free of such arbitrage. This shows that our two-stage training successfully provides a structural prior that guides the model to produce outputs that adhere to fundamental financial principles.

\begin{figure}[ht]
    \centering
    \includegraphics[width=\textwidth]{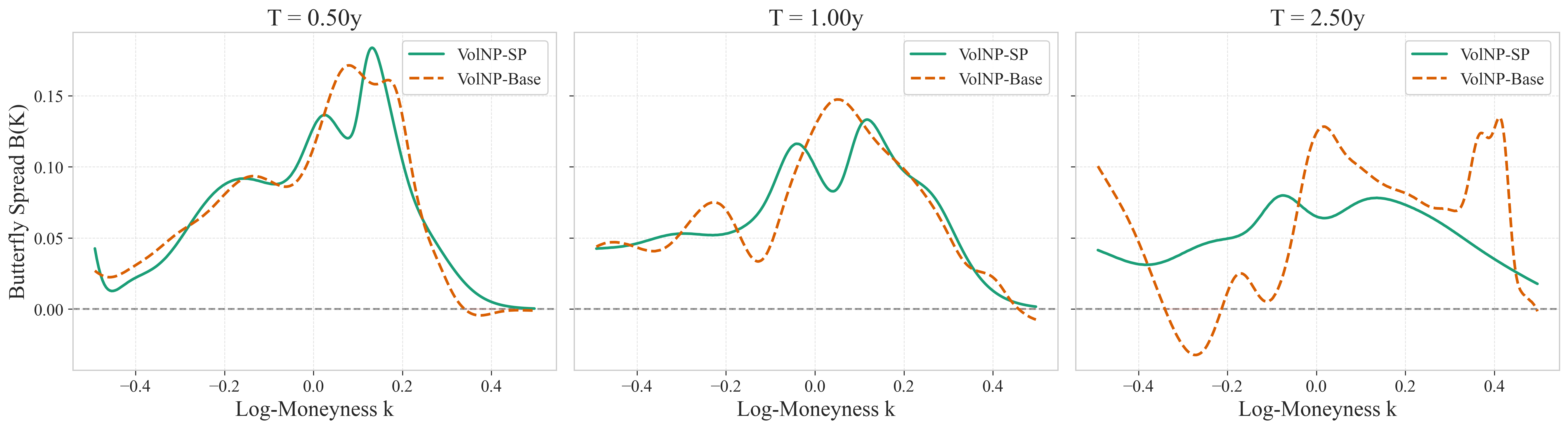}
    \caption{Butterfly-spread diagnostics on March 9, 2020. For $\tau=0.50$, $1.00$ and $2.50$, we compute discrete second differences of call prices on a uniform strike grid using each model's IV. Green curve is the result of VolNP-SP. Orange curve is the result of VolNP-Base.}
    \label{fig:butterfly_spreads}
\end{figure}

We further visualize two arbitrage diagnostics. First, for fixed maturities $\tau\in\{0.50,1.00,2.50\}$ we evaluate Black-Scholes call prices $C$ on a uniform strike grid around the forward and compute the discrete second difference $B(K)=C(K+h,\tau)-2C(K,\tau)+C(K-h,\tau)$ as a proxy for convexity in $K$. Nonnegative $B(K)$ rules out butterfly arbitrage. Figure~\ref{fig:butterfly_spreads} overlays the resulting curves. VolNP-SP yields smooth and largely nonnegative spreads across strikes, while the VolNP-Base shows negative values and oscillations on the wings, particularly at longer maturities.

\begin{figure}[ht]
    \centering
    \includegraphics[width=\textwidth]{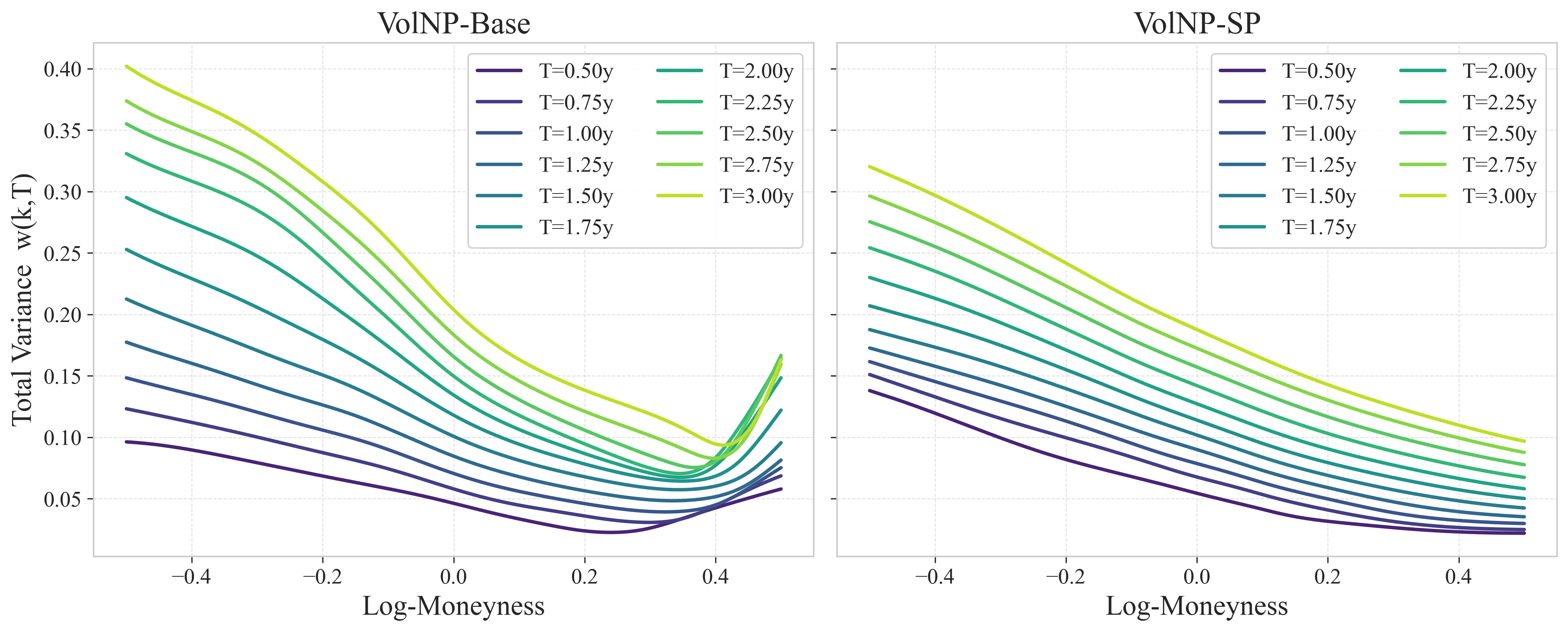}
    \caption{Total variance $w(k,\tau)$ on March 9, 2020. Curves show $w(\cdot,\tau)$ over log-moneyness for maturities from 0.50 year to 3.00 year. }
    \label{fig:total_variance}
\end{figure}

Second, we examine the total variance $w(k,\tau)=\sigma^2(k,\tau)\,\tau$. In a calendar-arbitrage-free surface, $w(k,\tau)$ should rise with $\tau$ at fixed log-moneyness $k$ and vary smoothly across $k$. Figure~\ref{fig:total_variance} shows families of $w(\cdot,\tau)$ curves. VolNP-SP exhibits nested, coherent term structures with well-behaved wings. VolNP-Base develops local humps at extreme $k$ that may compromise monotonicity. These views highlight the regularizing role of the SABR-induced prior under market stress.

\section{Conclusion}
\label{sec6}
This paper explores an alternative approach by framing surface construction as a meta-learning task. We propose the Volatility Neural Process, a model designed to learn a general mapping from sparse option quotes to a complete implied volatility surface. The model is first pre-trained on SABR-generated data to learn a financial-theoretic prior, and subsequently fine-tuned on market data to capture real-world dynamics. By learning the general task of surface construction, this framework yields a single model that does not require daily re-calibration, offering a potentially more efficient and reliable solution for constructing the implied volatility surface.

Empirical results on the SPX options dataset indicate that this approach is effective. The proposed model outperforms the considered benchmarks in overall accuracy and demonstrates robustness to data sparsity. The analysis also confirms that the pre-training stage is crucial for reducing large errors.

\textit{Future work.} The prior should be viewed as a modeling choice rather than a fixed specification. SABR is merely one example. Alternative priors can be derived from structural models (e.g., Heston, rough volatility) via synthetic surface generation, or from arbitrage-free parameterizations (e.g., SSVI). Future work can compare and combine priors through mixture or ensemble schemes and develop regime-adaptive procedures for prior selection.

\bibliographystyle{elsarticle-harv} 
\bibliography{references}

\end{document}